\documentstyle[pre,aps,multicol,epsfig,epsf]{revtex}
\renewcommand{\narrowtext}{\begin{multicols}{2} \global\columnwidth20.5pc}
\renewcommand{\widetext}{\end{multicols} \global\columnwidth42.5pc}
\newcommand{\Rrule}{\vspace{-0.1in}\hfill\vrule depth1em height0pt \vrule
  width3.5in height.2pt depth.2pt\vspace*{-0.125in}}
\newcommand{\Lrule}{\vspace*{-0.2in}\noindent\vrule width3.5in height.2pt
  depth.2pt \vrule depth0em height1em}

\begin{document}
\draft
\title{Spectra of massive and massless QCD Dirac operators: A novel link}
\author{G. Akemann$^1$ and E. Kanzieper$^2$}
\address{$^1$Max--Planck--Institut f\"ur Kernphysik, Saupfercheckweg 1,
  D-69117 Heidelberg, Germany \\
$^2$The Abdus Salam International Centre for Theoretical Physics, 
P.O. Box 586, I-34100 Trieste, Italy}
\date{31 January 2000}
\maketitle

\begin{abstract}
We show that integrable structure of chiral random matrix models
incorporating global symmetries of QCD Dirac operators 
(labeled by the Dyson index
$\beta=1,2$, and $4$) leads to emergence of a connection relation between the
spectral statistics of massive and massless Dirac operators. This novel link
established for $\beta$--fold degenerate massive fermions 
is used to explicitly derive (and prove the random matrix universality 
of) statistics of low--lying eigenvalues of
QCD Dirac operators in the presence of ${\rm SU(2)}$ massive fermions in the
fundamental representation ($\beta=1$) and ${\rm SU(} N_c \ge 2 {\rm )}$
massive adjoint fermions ($\beta=4$). Comparison with available
lattice data for ${\rm SU(} 2{\rm )}$ dynamical staggered fermions reveals 
a good agreement.
\end{abstract}


\narrowtext

Explicit knowledge of spectral statistics of low--lying 
eigenvalues of the Dirac operator is required to understand
the phenomenon of chiral symmetry breaking ($\chi$SB) in quantum chromodynamics 
(QCD). It has first been conjectured by Verbaarschot and collaborators 
\cite{V-1993} that extreme infrared limit of the QCD Dirac operator spectrum 
can be described by  the large--$N$ chiral Random Matrix Theory (RMT) that models 
the true Dirac operator ${\cal D}$ by $N\times N$ block offdiagonal matrix 
${\cal D}^{\rm RMT} = {\rm offdiag}(iW,iW^{\dagger})$, $W$ being $n\times m$ 
rectangular random matrix [see Eq. (\ref{jpdf}) below]. In such a formulation, $N=n+m$ 
is an analog of 
dimensionless space--time volume $V$, while $\nu=|n-m|$ is equivalent to the topological
charge (equal to the number of zero modes of ${\cal D}$). If, in addition, the entries 
of $W$ are chosen to be real, complex, or 
quaternion real, the random matrix ${\cal D}^{\rm RMT}$ possesses proper 
antiunitary symmetry (labeled by the Dyson index $\beta=1,2$, or $4$) and, hence, correctly 
reproduces both the underlying symmetries of the Dirac operator and the $\chi$SB
pattern associated with it. On the language of the chiral QCD Lagrangian, the above 
approach corresponds to the limit \cite{LS-1992} $1/\Lambda \ll V^{1/4} \ll 1/m_\pi$
($\Lambda$ is a typical hadronic scale and $m_\pi$ is the pion mass) in which the
kinetic term in Lagrangian can be neglected, and only the global symmetries 
of the Dirac operator become important. Recently, RMT phenomenology 
has been put onto a firm field theoretic 
ground represented by the framework of finite--volume partition functions \cite{AD-1} 
and the partially quenched perturbation theory \cite{OTV-1}.

On the microscopic scale $\sim 1/V\Sigma$, chiral RMT (defined for a given topological
charge $\nu$ \cite{N-1999}) leads to parameter--free predictions for the unfolded microscopic spectral 
density 
$\rho_S(\lambda) = \lim_{V\rightarrow \infty} (V\Sigma)^{-1} \rho(\lambda/V\Sigma)$
of the Dirac operator. Here, the absolute value $\Sigma$ of the chiral 
condensate (the order parameter of $\chi$SB) is related to the Dirac spectral density 
$\rho(0)$ at zero virtuality through the Banks--Casher relation $\Sigma = \pi \rho (0)/V$
\cite{BC-1980}. In a series of papers \cite{lattice}, RMT predictions have been
confronted to the lattice data, and good agreement has been found for the spectral 
density, two--level correlation function, and distribution of the smallest Dirac 
eigenvalue for {\it massless} lattice data of all $\chi$SB patterns. The lattice data for 
{\it massive} fermions have recently appeared \cite{BBMW-1998} as well.

Unfortunately, available theoretical results for microscopic spectral
correlators of
massive QCD Dirac operators are rather
poor, being restricted to the $\beta=2$ symmetry class 
\cite{JNZ-1996,DN-1998} associated
with the gauge group ${\rm SU(}N_c \ge 3{\rm)}$ in the fundamental    
representation. It
is the aim of the present Letter to 
show that 
integrable 
structure of chiral 
RMT
results in a simple but
powerful link between 
the spectral statistics of
massive and massless QCD Dirac operators for all three symmetry
classes  $\beta=1,2$, and
$4$. The connection relation 
[Eq. (\ref{Br})],
established below for 
$\beta$--fold degenerate 
massive fermions within the framework of chiral RMT,
relates partially unknown massive 
spectral correlation functions to the massless ones (taken at both
positive  and fictitious 
negative energies) \cite{K-1999}. As the 
latters
have already 
received a detailed 
study in the literature, this link
not only solves the problem posed but also provides a particularly simple
proof of RMT--universality of massive correlation functions, that becomes
a consequence of celebrated universality \cite{ADMN-1997,SV-1998,Rigor}
proven for the massless case.

Let us start with the definitions \cite{DN-1998,Mo-1991}. The joint probability distribution
function of
chiral random matrix ensemble associated with $N_f$ massive fermions
in the sector of
topological charge $\nu$ is given by
\begin{eqnarray}
\label{jpdf}
&&P_n^{(N_f,\nu ,\beta )}(\lambda_1,\ldots,\lambda_n ) = 
\frac{1} {Z_n^{(N_f,\nu ,\beta )}(\{m\})} \nonumber\\
&&\times \left| \Delta _n\left( \{ \lambda\} \right) \right|^\beta
\prod_{i=1}^n [ w_{\beta,\nu }(\lambda _i)
\prod_{f=1}^{N_f}m_f^\nu(\lambda _i+m_f^2)].
\end{eqnarray}
Here, $\{\lambda\} \ge 0$ are the eigenvalues of the matrix $WW^{\dagger}$,
$\Delta _n\left( \{\lambda\} \right)=\prod_{i<j}^n(\lambda_i-\lambda_j)$ 
is the Vandermonde determinant, the weight function $w_{\beta,\nu}$ is
$w_{\beta ,\nu }( \lambda ) =\lambda ^{\frac \beta
2\nu +\frac \beta 2-1}e^{- \beta V\left( \lambda \right)}$,
$V(\lambda)$ is the finite--polynomial confinement potential, and the topological 
charge $\nu$ is taken to be positive integer or zero. 

The $p$--point correlation function in the above ensemble is expressed as \cite{M-1991}
\begin{eqnarray}
\label{k-pt}
&&R_{n,p}^{(N_f,\nu ,\beta )}(\lambda _1,\ldots,\lambda _p) \ =  
\frac{n!}{(n-p)!} \\
&&\times \int_0^{+\infty }\!\!d\lambda _{p+1}\ldots d\lambda _n
P_n^{(N_f,\nu,\beta
  )}(\lambda_1,\ldots,\lambda_p,\lambda_{p+1},\ldots, \lambda_n). 
\nonumber
\end{eqnarray}
For $p=0$ this yields the mass--dependent partition function 
$Z_n ^{(N_f,\nu ,\beta )}(\{m\})$ appearing in Eq. (\ref{jpdf}).
The unfolded spectra of the Dirac operator 
are then obtained from the appropriately unfolded spectra 
${\hat R}_p^{(N_f,\nu,\beta)}(\{\lambda\})$ of associated random matrix 
model Eq. (\ref{jpdf})
by a simple transformation of variables:
\begin{eqnarray}
\label{spectra}
\rho_S(\lambda_1,\ldots,\lambda_p) = 2^p \prod_{k=1}^p |\lambda_k|
{\hat R}_p^{(N_f,\nu,\beta)}(\lambda_1^2,\ldots,\lambda_p^2).
\end{eqnarray}
We notice that for massive correlation functions this also demands to rescale
the quark masses, $\mu_f = m_f V\Sigma$. 
This completes our definition of the  model.

In what follows, we assume that the massive fermions are $\beta$--fold 
degenerate, 
${\cal M}_\beta=(m_1 \openone_\beta, \ldots, m_{N_f} \openone_\beta)$, so that
appropriate matrix ensemble is given by the joint probability distribution
function $P_n ^{(\beta N_f,\nu,\beta)}(\{\lambda\})$. Since in this case,
$P_n ^{(\beta N_f,\nu,\beta)}(\{\lambda\})$ contains a positive
definite  factor 
$\prod_{f=1}^{N_f}(\lambda_i+m_f^2)^\beta$, it can conveniently be
absorbed  into a 
Vandermonde determinant of a larger dimension 
\begin{eqnarray}
\label{delta}
\Delta _{n+N_f}(\{\lambda \},\{-m^2\})&\equiv& \Delta_n(\{\lambda \})
\Delta_{N_f}(\{-m^2\}) \nonumber\\
&\times&\prod_{i=1}^n\prod_{f=1}^{N_f}(\lambda_i+m_f^2). 
\end{eqnarray}
This immediately results in a pretty fact that the partition function of
the model Eq. (\ref{jpdf}) with $\beta N_f$ massive fermions can be expressed
in terms of associated {\it massless} $N_f$--point correlation function
$R_{n+N_f,N_f}^{(0,\nu,\beta)}(\{-m^2\})$ of the matrix ensemble of larger dimension, 
$(n+N_f)\times(n+N_f)$, taken at fictitious {\it negative} energies: 
\begin{eqnarray}
\label{Z} 
\frac{Z_n^{(\beta N_f,\nu ,\beta )}(\{m\})}{Z_{n+N_f}^{(0,\nu ,\beta )}}&=& 
\frac{n!}{(n+N_f)!}
\left( \prod_{f=1}^{N_f}\frac{m_f^{\beta \nu }}{w_{\beta,\nu }(-m_f^2)}\right) 
\nonumber\\ 
&\times&
\frac{R_{n+N_f,N_f}^{(0,\nu ,\beta)}(-m_1^2,\ldots,-m_{N_f}^2)}
{\left| \Delta_{N_f}(\{-m^2\})\right|^\beta}.
\end{eqnarray}
The same strategy is applied to the $p$--point correlator, Eq. (\ref{k-pt}). 
After a few transformations,
we arrive at the following remarkable relationship:
\begin{eqnarray}
\label{Br}
&&R_{n,p}^{(\beta N_f,\nu ,\beta )}(\lambda _1,\ldots,\lambda _p) \nonumber\\
&&=\frac{R_{n+N_f,p+N_f}^{(0,\nu ,\beta )}(\lambda _1,\ldots,\lambda_p,
-m_1^2,\ldots,-m_{N_f}^2)}{R_{n+N_f,N_f}^{(0,\nu ,\beta)}
(-m_1^2,\ldots,-m_{N_f}^2)}.
\end{eqnarray}
Finite--$n$ Eq. (\ref{Br}) establishes a link between massive and
massless  spectral 
correlators via associated chiral random matrix ensemble, and
represents a  basic relation 
to be examined in the rest of the Letter, where we consider the most interesting
symmetry classes $\beta=1$ and $4$.
It should be stressed that, in spite of seeming
simplicity, the link Eq. (\ref{Br}) is not obvious as it involves correlation
functions of massless chiral ensemble taken at both positive and fictitious
{\it negative} energies. 
We also wish to emphasize that, in the microscopic
limit, Eq. (\ref{Br})
immediately leads to the RMT--universality of the microscopic massive correlators
which becomes a simple consequence of the universality phenomenon established
for massless correlation functions \cite{ADMN-1997,SV-1998}.

(i) Let us turn to the $\beta=4$ symmetry class, associated with
${\rm SU(}N_c \ge 2{\rm )}$ massive adjoint fermions. In the 
vicinity of the hard edge, the unfolded $p$--point spectral correlators in the 
{\it massless} chiral 
model Eq. (\ref{jpdf}), $N_f=0$, admit 
quaternion determinant representation \cite{M-1991}
\begin{eqnarray}
\label{pf4}
{\hat R}_p ^{(0,\nu,4)}(\{\lambda\}) &=& {\rm Qdet} \left[ f_4 (\lambda_i,\lambda_j) 
\right]_{1\le i,j \le p}.
\end{eqnarray}
Here, the $2 \times 2$ matrix kernel $f_4\equiv f_{\beta=4}$ \cite{TW-1998}
\widetext
\Lrule
\begin{eqnarray}
\label{f4}
f_\beta(X,Y) = 
\left(
\begin{array}{cc} 
S_\beta (X,Y)
& 
D_\beta(X,Y)
\\ 
I_\beta(X,Y)
& 
S_\beta(Y,X)
\end{array} 
\right),\;
D_\beta(X,Y) = -\partial_Y S_\beta(X,Y), \;
I_\beta(X,Y) = \int_Y ^X dZ S_\beta(Z,Y)-\epsilon(X-Y)\delta_{\beta,1},
\end{eqnarray}
\Rrule
\narrowtext
\noindent
$\epsilon(X)=(1/2){\rm sgn}(X)$, is uniquely specified by 
the function 
\cite{FNH-1999}
\begin{eqnarray}
\label{S4}
S_4(X,Y) &=& 2K_{2\nu+1}(2X^{1/2},2Y^{1/2}) \nonumber \\
&-& \frac{J_{2\nu}(2X^{1/2})}{4X^{1/2}}
\int_0^{2Y^{1/2}}\!\!dtJ_{2\nu+2}(t)
\end{eqnarray}
with $K_\alpha(X,Y)$ being the Bessel kernel \cite{TW-1994}:
\begin{equation}
\label{ker-2}
K_\alpha (X,Y) = 
\frac{XJ_{\alpha+1}(X)J_\alpha(Y)-Y J_{\alpha+1}(Y)
J_\alpha(X)}{2(X^2-Y^2)}.
\end{equation}
It is important to stress that Eqs. (\ref{S4}) and (\ref{ker-2}) hold generically
\cite{SV-1998} for arbitrary finite--polynomial confinement potential $V(\lambda)$ in 
Eq. (\ref{jpdf}). 

As long as at $\beta=4$ the definitions Eqs. (\ref{jpdf}) and (\ref{k-pt})
do not discriminate between the positive and (fictitious) negative eigenvalues 
$\lambda_k$, Eqs. (\ref{S4}) and (\ref{ker-2}) remain valid
at negative arguments also, 
provided one makes use of the fact that $J_\alpha (iX) = i^\alpha I_\alpha(X)$. This
circumstance allows us, at once, to derive closed expressions for $p$--point 
{\it massive} correlation functions in the microscopic limit.
Straightforward calculations based on Eqs. (\ref{spectra}), (\ref{Br}) and (\ref{pf4}) 
yield:

\widetext
\begin{figure}[-b]
\centerline{\epsfig{figure=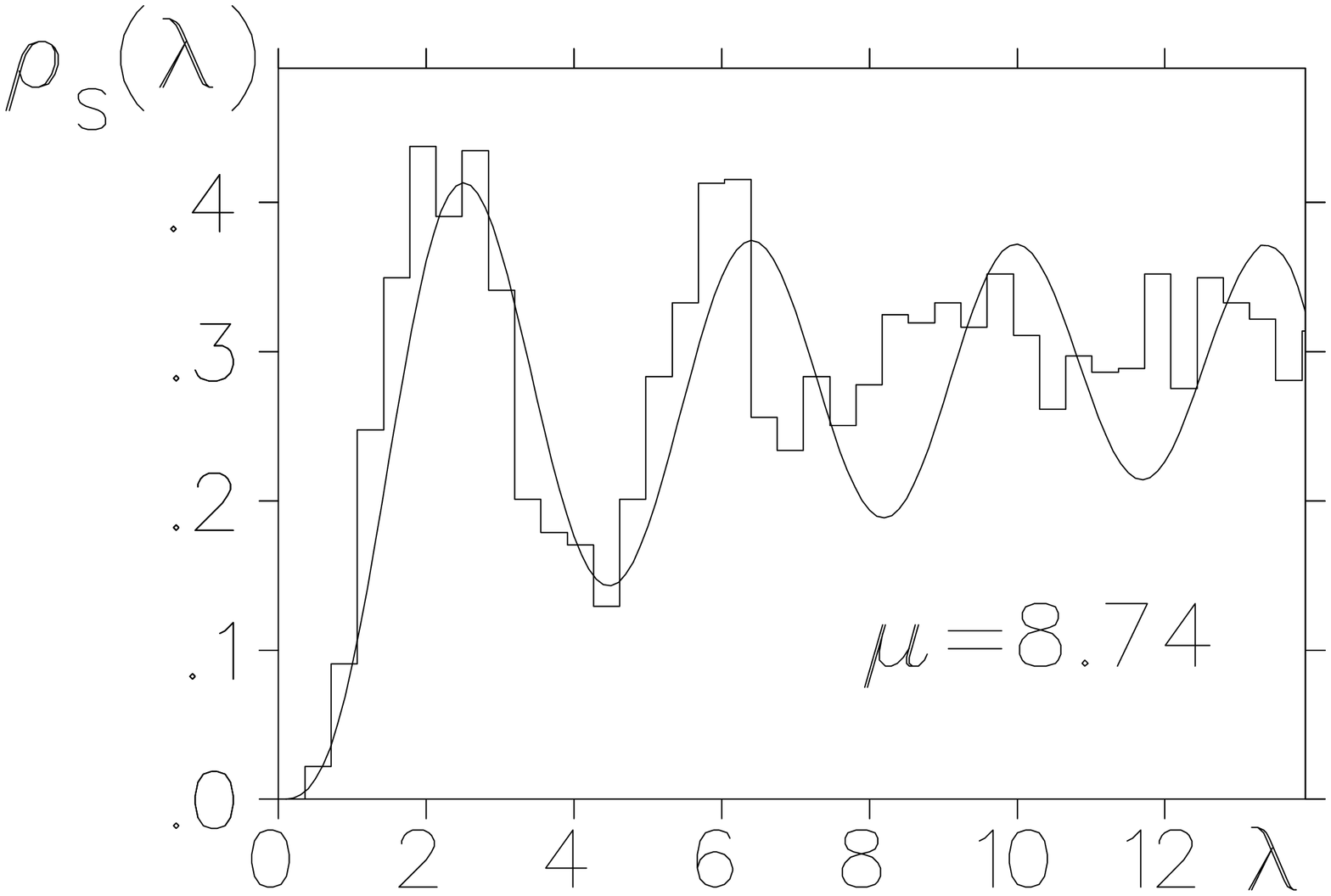,width=9pc,angle=90}
\epsfig{figure=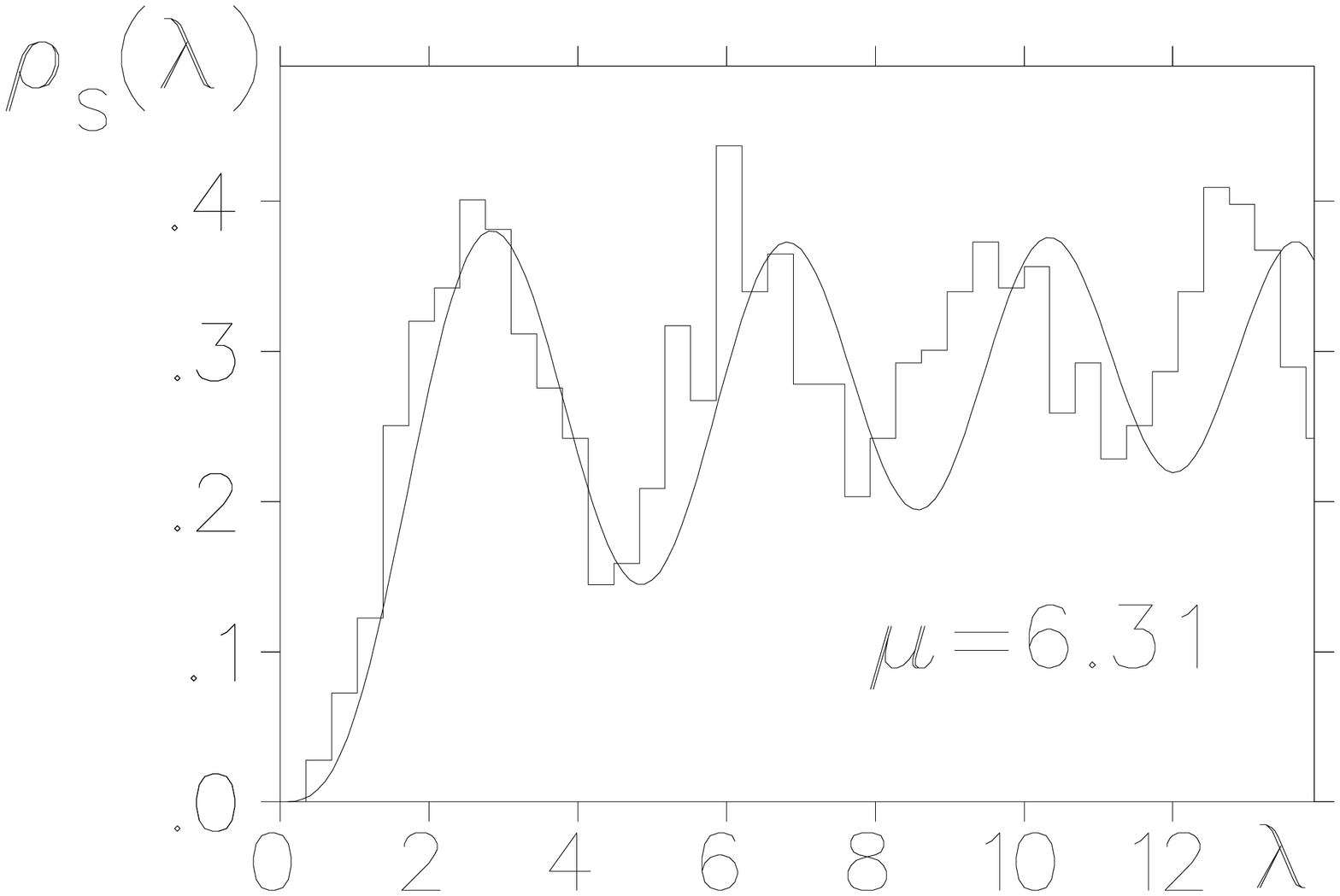,width=9pc,angle=90}
\epsfig{figure=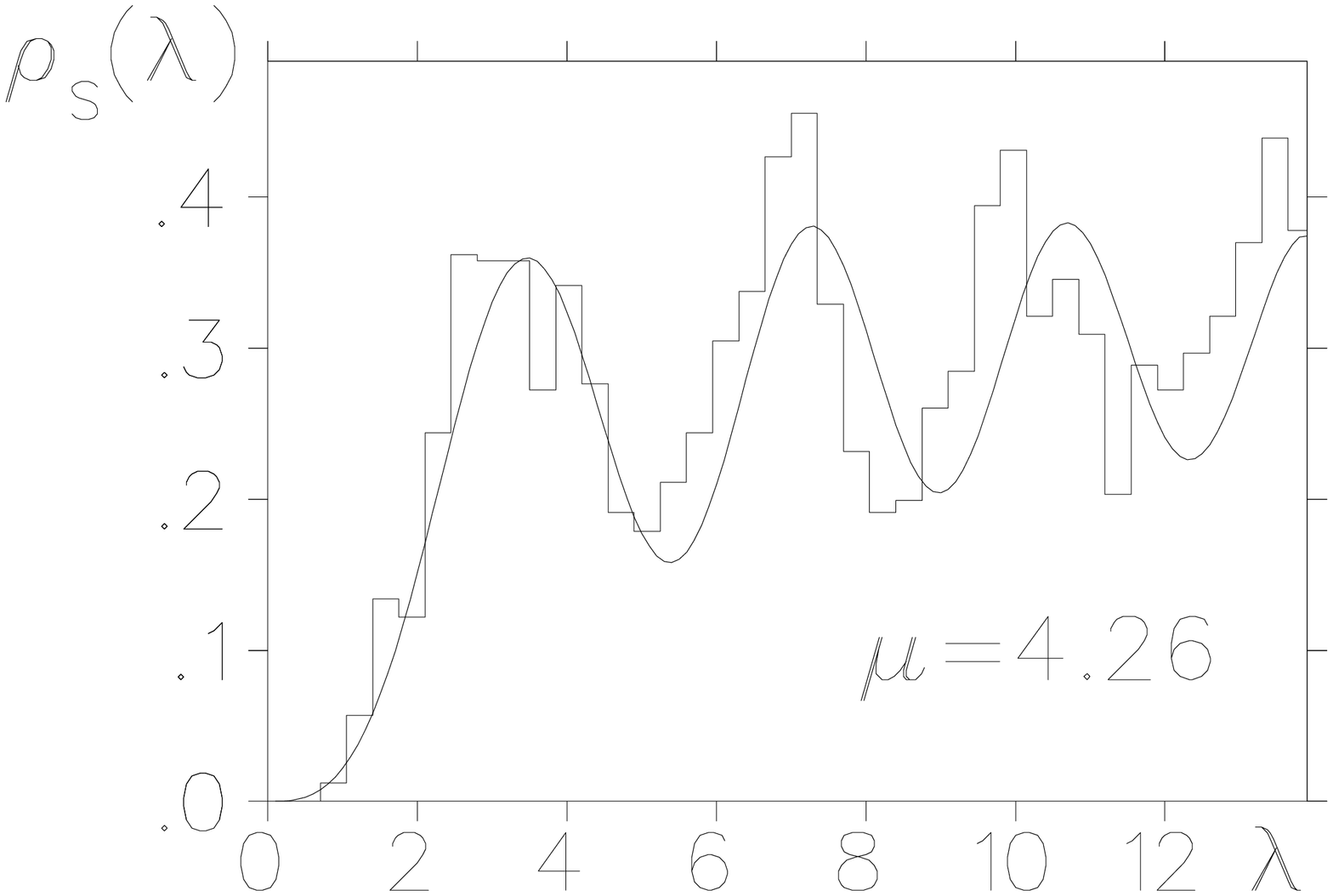,width=9pc,angle=90}}
\caption{
Microscopic massive density $\rho_S^{(4)}(\lambda;\mu \openone_4)$, Eq. (12), plotted
versus lattice data simulated in Ref. [8] for 8 flavors of masses $\mu=8.74,6.31,$
and $4.26$. 
For discussion of species doubling and systematic deviations at large $\lambda$ we refer 
to Ref. [8].
}
\label{fig1}
\end{figure}
\narrowtext
\noindent

\begin{eqnarray}
\label{cf-f4}
\rho_S^{(4N_f)} (\lambda_1,\ldots,\lambda_p;\mu_1\openone_4,\ldots,\mu_{N_f}\openone_4)
= 2^p \prod_{k=1}^p |\lambda_k| \nonumber \\
\times
\frac
{
{\rm Qdet}
\left[
\begin{array}{cc} 
f_4 (\lambda_i^2,\lambda_j^2)
& 
f_4(\lambda_i^2,-\mu_{f^\prime}^2)
\\ 
f_4(-\mu_f^2,\lambda_j^2)
& 
f_4(-\mu_f^2,-\mu_{f^\prime}^2)
\end{array}
\right]
}
{
{\rm Qdet}
[f_4(-\mu_f^2,-\mu_{f^\prime}^2)]
},
\end{eqnarray} 
$1\le i,j \le p$, $1\le f,f^\prime \le N_f$. 
We remind that this result applies to 
$4$--fold degenerate quark masses, $\mu_f=m_f V\Sigma$. 

Microscopic density for the particular case of 4 degenerate fermions of mass $\mu$ 
is of special interest as it can be compared to available lattice data for
dynamical ${\rm SU(}2{\rm)}$ staggered fermions in the fundamental representation 
simulated in Ref. \cite{BBMW-1998} at $\nu=0$. [Because of the lattice symmetry 
of staggered fermions they belong to the symmetry class $\beta=4$]. Computing quaternion
determinants in Eq. (\ref{cf-f4}), we come down to
\widetext
\Lrule
\begin{eqnarray}
\label{4mass}
\rho_S^{(4)}(\lambda;\mu\openone_4) = 2|\lambda| \left( S_4(\lambda^2,\lambda^2)
-\frac
{
S_4(-\mu^2,\lambda^2)S_4(\lambda^2,-\mu^2) - I_4(\lambda^2,-\mu^2)D_4(\lambda^2,-\mu^2)
}
{
S_4(-\mu^2,-\mu^2)
}
\right).
\end{eqnarray}
\Rrule
\narrowtext
\noindent
We have explicitly checked that the limit $\mu\rightarrow 0$ reproduces the known
massless result \cite{V-1993} at a shifted topological charge $\nu\rightarrow \nu+2$
\cite{remark}. Theoretical results plotted in Fig. 1 for three different values 
of $\mu$ show reasonable agreement with numerical data.

We close our consideration of $\beta=4$ symmetry class by giving a compact expression for
the 
previously unknown
massive RMT (or finite--volume \cite{HV-1995}) partition function
with $4$--fold degenerate massive fermions [see Eq. (\ref{Z})]:

\begin{equation}
\label{Z4}
\tilde{Z}^{(4N_f)}_\nu(\mu_1 \openone_4,\ldots,\mu_{N_f} \openone_4)\! = 
\!\frac{(-1)^{N_f}{\rm Qdet}[f_4 (-\mu_f ^2,-\mu_{f^\prime}^2)]}
{|\Delta_{N_f}(-\mu_f^2)|^4 \prod_{f=1}^{N_f}\mu_f^2}.
\end{equation}
Here, only nontrivial mass dependence has been displayed.

(ii) Now we turn to the $\beta=1$ symmetry class associated with ${\rm SU(}2{\rm )}$
massive fermions in the fundamental representation. In this case,
the modulus of the Vandermonde determinant in Eqs. (\ref{jpdf})
and (\ref{k-pt})
makes all $p$--point correlation functions to be nonanalytic 
functions of their arguments. This is exactly the reason of why one 
cannot use known expressions for massless correlation functions
${\hat R}_p^{(0,\nu,1)}(\{\lambda\})$ to naively compute them at
negative energies. Below we show
how to circumvent this obstacle for the simplest situation of the
spectral  density with a single quark mass. Extension to higher order correlation
functions  and/or 
larger number of masses is straightforward. 

In accordance with the connection relation Eq. (\ref{Br}), the
finite--$n$  massive spectral density equals
\begin{eqnarray}
\label{dos-1}
R^{(1,\nu,1)}_{2n-1,1}(\lambda) = \frac{R_{2n,2}^{(0,\nu,1)}(\lambda,-m^2)}
{R_{2n,1}^{(0,\nu,1)}(-m^2)},
\end{eqnarray}
where, for definiteness, we have fixed the dimension of the
massless  random matrix
ensemble to be even, $2n$; from now on, the 
superscripts 
are omitted
for
brevity. 
We observe that the
function $R_{2n,2}(\lambda,-m^2)$ can be
evaluated  through
the functional derivative of $R_{2n,1}(-m^2;[W])\equiv R_{2n,1}(-m^2)$ with 
respect to the 
confinement
potential $W$, $\exp\{-W(\lambda)\}=\lambda^{(\nu-1)/2}\exp\{-V(\lambda)\}$:
\begin{eqnarray}
\label{fd}
R_{2n,2}(\lambda,-m^2) &\equiv& R_{2n,1}(-m^2)\left(
R_{2n,1}(\lambda) -\delta(\lambda+m^2)\right) \nonumber \\
&-& \frac{\delta}{\delta W(\lambda)} R_{2n,1}(-m^2;[W]).
\end{eqnarray}
To facilitate taking the
functional  derivative in Eq. (\ref{fd}), we
utilize the  approach of
Ref. \cite{NF-1995} [see Eq. (A6) of second reference], but
express $R_{2n,1}(-m^2;[W])$ in terms of arbitrary polynomials $p_j(x)$
rather than
in terms of the skew orthogonal ones:
\begin{eqnarray}
\label{tw}
R_{2n,1}(-m^2;[W]) &=& \frac{1}{2} e^{-W(-m^2)} \sum_{j,k=0}^{2n-1} p_j(-m^2)
\mu_{jk}[W] \nonumber\\
&&\times \int_0^{+\infty} \!\!dZ\ e^{-W(Z)}p_k(Z).
\end{eqnarray}
The $2n \times 2n$ real antisymmetric matrix $\mu_{jk}[W]$ is the
inverse to  the
matrix \cite{TW-1998}
\begin{equation}
\label{M-matrix}
M_{jk}= \int_0^{+\infty} \!\!dx dy\  e^{-W(x)-W(y)}\epsilon(x-y)
p_j(x)  p_k(y),
\end{equation}
$\epsilon(x)=(1/2){\rm sgn}(x)$. Substituting Eq. (\ref{tw}) into
Eq.  (\ref{fd}),
and then into Eq. (\ref{dos-1}), we are able to express the finite--$n$ 
massive spectral
density $R_{2n-1,1}^{(1,\nu,1)}(\lambda)$ in the form
\begin{eqnarray}
\label{mass-dos}
&&R_{2n-1,1}^{(1,\nu,1)}(\lambda) \ =\ S_1^{(2n)}(\lambda,\lambda) \\
&&-\frac{
S_1^{(2n)}(-m^2,\lambda)S_1^{(2n)}(\lambda,0) - I_1^{(2n)}(0,\lambda)
D_1^{(2n)}(-m^2,\lambda)
}
{
S_1^{(2n)}(-m^2,0)
}\nonumber
\end{eqnarray}
that contains the entries of finite--$n$, $2 \times 2$ matrix kernel
$f_1^{(2n)}(X,Y)\equiv f_{\beta=1}^{(2n)}(X,Y)$ of the massless ensemble 
[see Eq. (\ref{f4})].
In deriving Eq. (\ref{mass-dos}) we have used both the representation \cite{TW-1998} 
$S_1^{(2n)}(X,Y)=-e^{-W(X)}\sum_{j,k=0}^{2n-1} p_j(X)\mu_{jk} 
\int_0^{+\infty}dZ \epsilon(Y-Z)e^{-W(Z)}p_k(Z)$, and Eq. (\ref{f4}).
Finally, taking into account Eqs. (\ref{spectra}), 
(\ref{mass-dos}), and the universal \cite{SV-1998} formula \cite{FNH-1999} 
\begin{eqnarray}
\label{S1-univ}
S_1(X,Y) &=& K_{\nu-1}(X^{1/2},Y^{1/2}) \nonumber\\
&-& \frac{J_{\nu}(X^{1/2})}{4X^{1/2}}
\left(\int_0^{Y^{1/2}}\!\!dtJ_{\nu-2}(t)\ -\ 1\right)
\end{eqnarray}
(valid in the vicinity of the hard edge), we deduce a closed expression for the 
microscopic single--mass spectral density $\rho_S^{(1)}(\lambda;\mu)$. It exhibits the 
quaternion determinant structure of Eq. (\ref{4mass}) with obvious changes in
arguments of $S$, $D$, and $I$ functions as is given by Eq. (\ref{mass-dos}). As 
a consistency check, we have verified that for $\mu=0$ it reduces to the
known result \cite{V-1993} for one massless flavor. Let us stress, 
that universal form \cite{SV-1998} of the function $S_1$ thus confirms the universality of 
the massive spectral density following in a more general context directly from the
microscopic limit of the connection relation Eq. (\ref{Br}).

In conclusion, we have derived universal expressions for 
spectral correlators
of massive chiral 
matrix ensembles
corresponding to $\beta$--fold degenerate massive fermions, by establishing
a new link between the 
statistics of massive and massless random matrices. 
The results obtained have been compared to the available lattice data associated
with $\beta=4$ $\chi$SB pattern in low--energy QCD.

We thank the authors of Ref. \cite{BBMW-1998} for making their data available to us. 
E.K. acknowledges early discussion with S.M. Nishigaki. The work of G.A. was 
partially supported by EU--TMR grant No. ERBFMRXCT97--0122. 

Note added.--After completing this work, the preprint \cite{NN-2000} by T. Nagao and 
S.M. Nishigaki
on finite--volume partition functions has appeared. In particular, these 
authors give alternative representations of our Eqs. (\ref{cf-f4}) and (\ref{Z4}).

\widetext

\vspace{-0.6cm}
\begin{references}
\vspace{-1.6cm}

\bibitem{V-1993} E.V. Shuryak and J.J.M. Verbaarschot,
Nucl. Phys. A  {\bf 560}, 306 (1993); J.J.M. Verbaarschot and I. Zahed, 
Phys. Rev. Lett. {\bf 70}, 3852 (1993); J.J.M. Verbaarschot, ibid. 
{\bf 72}, 2531 (1994).

\bibitem{LS-1992} H. Leutwyler and A. Smilga, Phys. Rev. D {\bf 46},
5607  (1992); J. Gasser and H. Leutwyler, Phys. Lett. B {\bf 188},
477  (1987); Nucl. Phys. B {\bf 307}, 763 (1988).

\bibitem{AD-1} P.H. Damgaard, Phys. Lett. B {\bf 424}, 322 (1998); G.
  Akemann and P.H. Damgaard, ibid. B {\bf 432}, 390 (1998); Nucl. Phys. B 
{\bf 519}, 682 (1998).

\bibitem{OTV-1}
J.C. Osborn, D. Toublan, and J.J.M. Verbaarschot,
Nucl.  Phys. B
{\bf 540}, 317 (1999); P.H. Damgaard, J.C. Osborn, D. Toublan, and J.J.M. 
Verbaarschot, ibid. B {\bf 547}, 305 (1999); D. Toublan and
J.J.M. Verbaarschot, ibid. B {\bf 560}, 259 (1999).

\bibitem{N-1999} F. Niedermayer, Nucl. Phys. Proc. Suppl. {\bf 73},
105 (1999) and references therein for topology on the lattice.


\bibitem{BC-1980} T. Banks and A. Casher,  Nucl. Phys. B {\bf 169}, 103 (1980).

\bibitem{lattice} 
M.E. Berbenni-Bitsch, S. Meyer, A. Sch\"afer, J.J.M. 
Verbaarschot, and T. Wettig, Phys. Rev. Lett. {\bf 80}, 1146 (1998);
R.G. Edwards, U.M. Heller, J. Kiskis,  and R. 
Narayanan, ibid. {\bf 82}, 4188 (1999); 
M. G\"ockeler, H. Hehl, P.E.L. Rakow, A. Sch\"afer, and T. Wettig, 
Phys. Rev. D {\bf 59}, 94503 (1999); R.G. Edwards,
U.M.  Heller, and R. Narayanan, ibid. D {\bf 60}, 77502 (1999);
P.H. Damgaard, U.M. Heller, and A. Krasnitz, 
Phys. Lett. B {\bf 445}, 366 (1999). 

\bibitem{BBMW-1998} M.E. Berbenni-Bitsch, S. Meyer, and T. Wettig,
Phys. Rev. D {\bf 58}, 71502 (1998).

\bibitem{JNZ-1996} J. Jurkiewicz, M.A. Nowak, and I. Zahed,
Nucl. Phys. B {\bf 478}, 605 (1996); Erratum: ibid. B {\bf 513}, 759 (1998).

\bibitem{DN-1998} P.H. Damgaard and S.M. Nishigaki, Nucl. Phys. B
  {\bf 518}, 495 (1998); S.M. Nishigaki, P.H. Damgaard, and T. Wettig, 
Phys. Rev. D {\bf 58}, 87704  (1998); T. Wilke, T. Guhr, and T. Wettig, 
ibid. D {\bf 57}, 6486 (1998).


\bibitem{K-1999} A conceptually similar idea has recently been
realized to prove existence of a connection relation between parametric and
conventional level 
statistics in non--Gaussian RMT: E. Kanzieper, Phys. Rev. Lett. {\bf
  82}, 3030  (1999).


\bibitem{ADMN-1997} G. Akemann, P.H. Damgaard, U. Magnea, and S. Nishigaki,
Nucl. Phys. B {\bf 487}, 721 (1997); E. Kanzieper and V. Freilikher,
Philos.  Magazine B {\bf 77}, 1161 (1998).

\bibitem{SV-1998}M.K. \c{S}ener and J.J.M. Verbaarschot,
Phys. Rev. Lett. {\bf 81}, 248 (1998); H. Widom, J. Stat. Phys. 
{\bf 94}, 347 (1999).

\bibitem{Rigor} V. Freilikher, E. Kanzieper, and I. Yurkevich, Phys. Rev. E
{\bf 53}, 2200 (1996); ibid. E {\bf 54}, 210 (1996); P. Bleher and A. Its, Ann. of 
Math. {\bf 150}, 185 (1999); P. Deift, T. Kriecherbauer, K.T.-R. McLaughlin, 
S. Venakides, and X. Zhou, Commun. Pure Appl. Math. {\bf 52}, 1335 (1999).

\bibitem{Mo-1991} T.R. Morris, Nucl. Phys. B {\bf 356}, 703 (1991).

\bibitem{M-1991} M.L. Mehta, {\it Random Matrices} (Academic, San
  Diego, 1991).

\bibitem{TW-1998} C.A. Tracy and H. Widom, J. Stat. Phys. {\bf 92},
  809 (1998).

\bibitem{FNH-1999} P.J. Forrester, T. Nagao, and G. Honner,
Nucl. Phys. B {\bf 553}, 601 (1999).

\bibitem{TW-1994} C.A. Tracy and H. Widom, Commun. Math. Phys. {\bf 161},
289 (1994).

\bibitem{remark} Indeed, as $\beta$--fold degenerate masses $m_f$ 
(of total amount of $\beta N_f$) approach zero, the
weight function $w_{\beta,\nu}$ in Eq. (\ref{jpdf}) acquires the factor 
$\lambda ^{\beta N_f}$. This is equivalent to the effective massless
measure $w_{\beta,\nu+2N_f}$ corresponding to the sector with a modified 
topological charge $\nu+2N_f$.

\bibitem{HV-1995} M.A. Halasz and J.J.M. Verbaarschot, 
Phys. Rev. D {\bf 52}, 2563 (1995).

\bibitem{NF-1995} T. Nagao and P.J. Forrester, Nucl. Phys. B {\bf 435},
401 (1995); ibid. B {\bf509}, 561 (1998).

\bibitem{NN-2000} T. Nagao and  S.M. Nishigaki, eprint hep--th/0001137.

\end{references}
\end{document}